\documentclass[journal=jacsat,manuscript=article]{achemso}
\usepackage[version=3]{mhchem}
\usepackage{amsfonts}
\usepackage{graphicx}
\usepackage{amssymb}
\usepackage{color}

\author{Yingteng Zhai}
\affiliation{Key Laboratory for Computational Physical
Sciences(MOE) and Surface Physics Laboratory $\&$ Department of Physics, Fudan
University, Shanghai 200433, China}
\author{Alessandro Laio}
\affiliation{International School for Advanced Studies (SISSA), via Bonomea 265, 34136 Trieste, Italy.}
\affiliation{CNR-IOM Democritos National Simulation Centre, via Bonomea 265, 34136 Trieste, Italy.}
\author{Erio Tosatti}
\affiliation{International School for Advanced Studies (SISSA), via Bonomea 265, 34136 Trieste, Italy.}
\affiliation{Abdus Salam International Centre for Theoretical Physics (ICTP), Strada Costiera 11, 34014 Trieste, Italy.}
\affiliation{CNR-IOM Democritos National Simulation Centre, via Bonomea 265, 34136 Trieste, Italy.}
\author{Xin-Gao Gong}
\email{xggong@fudan.edu.cn}
\affiliation{Key Laboratory for Computational Physical Sciences(MOE)
 and Surface Physics Laboratory $\&$ Department of Physics, Fudan University, Shanghai 200433, China}
\title{Finite temperature properties of clusters by replica exchange metadynamics: the water nonamer}
\begin{document}
\begin{abstract}
We introduce an approach for the accurate calculation of thermal properties of classical nanoclusters.
Based on a recently developed enhanced sampling technique, replica exchange metadynamics, the method yields the true free energy
of each relevant cluster structure, directly sampling its basin and measuring its occupancy in full equilibrium.
All entropy sources, whether vibrational, rotational anharmonic and especially configurational -- the latter often forgotten in many
cluster studies -- are automatically included. For the present demonstration we choose the water nonamer
\ce{(H2O)9}, an extremely simple cluster which nonetheless displays a sufficient complexity and interesting
physics in its relevant structure spectrum. Within a standard TIP4P potential description of water,
we find that the nonamer second relevant structure possesses a higher configurational entropy
than the first, so that the two free energies surprisingly cross for increasing temperature.
\end{abstract}

\section{Introduction}

The physics of clusters formed by a small number of atoms or molecules
is basic to many branches of science\cite{baletto_ferrando_2005}. Clusters are the primary nanosystem
where matter behaves differently from its usual bulk aggregation.
In condensed matter physics it has long been of interest how collective
features of macroscopic systems, such as ordering phenomena and phase
transitions, emerge in a nutshell in small size clusters\cite{wales_berry_1992_phase_trans,
proykova_berry_2006}.
As is well known, it often proves harder to predict the properties
of a small aggregate than those of an infinitely large system, where
one can take advantage of regularity. In computational physics,
identifying and characterizing cluster structures provide a prototypically
difficult optimization problem\cite{wales_doye_2000}.
The potential energy landscape as a function of the atomic or molecular
coordinates is generally complex, with many local minima separated
by large barriers even in relatively small clusters, where the simple
identification of the lowest energy structure may be a serious computational
challenge. State-of-the-art optimization techniques are for example
sometimes benchmarked on their capability to find the global energy
minimum of Lennard-Jones clusters\cite{berry_1987, wales_doye_1997, car_gregor_2005}.
{
Simulation studies have been devised,  using enhanced sampling   methods,
for example
aggregation-volume-bias Monte Carlo and umbrella sampling,
to obtain the free energy and density of state of clusters or crystalline nuclei
in liquids \cite{JPCC_Al_2007,vapor_liquid_water_2005,gas_liquid_1998,nano_olids_slush_2008}.
}
The recent boost of nanoscience increasingly demands a better and better ability to characterize,
calculate, and understand theoretically the peculiar properties of matter
at the nanoscale, a subject of broad technological, practical as well as conceptual
implications.

The traditional manner to approach cluster simulation is quite simple. As a first step
one attempts to identify all the locally stable low energy structures, usually
by some global optimization technique, such as simulated annealing, genetic algorithms,
etc . The computational cost of these techniques is largely dominated by repeated
potential energy evaluation, and one must carefully choose a level of description
of the mechanics of atomic coordinates (quantum or semiempirical or
classical) and of the electronic degrees of freedom allowing at the
same time an exhaustive search and a sufficiently accurate estimate
of the energy. Sometimes a preliminary search with a less accurate
empirical intermolecular potential can be followed by a subsequent
refinement leading to a more accurate energy estimate and thus to
the identification of the relevant structures with a higher level
description. If one is interested in the properties of the system
at a very low temperature, what matters is the global energy minimum.
When however temperature grows so that $k_{B}T$ is comparable to
the energy gap $\Delta_{01}$ above the global minimum, higher energy
states become relevant and must be identified. It is traditional to
consider a relatively small number of ``relevant states'', analogous
to inherent states in glass physics, crudely consisting of locally stable
atomic configurations or cluster geometries, that identify local energy
minima in the space of atomic coordinates. A finite set of relevant
states can be obtained for example by (real or conceptual) annealing
of the system from a much larger number and variety of finite temperature
instantaneous molecular configurations, cooled down to reach the closest
local energy minimum at $T=0$. At finite temperature, each relevant
state $s$ represents the center of a basin in configuration space.
The logarithm of the volume of the relevant state basin volume  in configuration
space represents its entropy $S_{s}$. When the thermal occupancy
of each relevant structure $s$ is explicitly calculable, this is
equivalent to specifying its free energy $F_{s}=E_{s}-TS_{s}$ . A
crucial question that needs to be addressed is, what contributes to
the entropy of the relevant state, and how it can be properly calculated.
The simplest and commonest approach is to consider {free rotations} and harmonic vibrations around
the $T=0$ local energy minimum of the relevant state, and calculate
the vibrational 
{
and rotational}
entropy involved, in the 
{
rigid-rotor-harmonic}
approximations\cite{jacs_Al_2007,JPCC_Al_2007,LJ38_2008}. 
This is however not generally adequate, and in fact
the rot-vibrational approximation may yield free energies of relevant
structures in serious error. The full entropy of a state, measuring
the size of its stability basin should include not just vibrations and rotations,
but all possible coordinate transformations leading to essentially
equivalent configurations. Proper determination of configurational
entropy is a factor which critically reflects the abundance of a relevant
state in the cluster's thermal equilibrium. Although these concepts
are natural, and well rooted in classic cluster literature
\cite{wales_berry_1992_phase_trans,proykova_berry_2006}, they are otherwise
not always properly implemented in everyday practice
where configurational entropy, difficult to define and evaluate, is
often overlooked. A practical and accurate calculation approach including automatically
all sources of entropy, especially for the more complex molecular clusters,
is highly desirable.

In this paper we introduce and demonstrate a procedure leading to
a computationally convenient and conceptually correct extraction
of the free energy of the relevant states of a cluster at finite temperature,
including all non-electronic entropy sources contributions. In the
simple case we use for our demonstration, a small water cluster, the
entropy is found to differ drastically between one relevant state
and another. This working case also shows explicitly that the neglect of
configurational contributions leads to large errors, totally unacceptable
practically no less than conceptually. Our method simultaneously addresses
both steps normally followed in traditional cluster study techniques:
a) the search for the relevant structures, and b) the study of their
finite-temperature properties. The approach, based on a recently developed
enhanced sampling technique, replica exchange metadynamics,
\cite{Metadynamics_method}
allows the computation of the true free energy of each relevant structure,
directly sampling its basin and measuring its occupancy (overall
abundance) in full thermal equilibrium, thus including automatically
all entropy sources.

For our demonstration purposes we chose the water nonamer \ce{(H2O)9},
a very simple dielectric cluster which nonetheless displays a sufficient complexity
and interesting physics in its relevant structure spectrum. Although
not directly addressed in this study, well characterized spectroscopic
properties of this particular cluster are available.\cite{buck2003}
We studied the cluster properties covering a temperature range of
potential interest for atmospheric science (100-250 K), and described
the forces between water molecules with a simple but realistic and
widely tested empirical force scheme, the so-called TIP4P potential\cite{TIP4P}.
Since the electronic shells of this system are closed and widely gapped,
the use of such effective interatomic forces is not unreasonable,
although clearly not definitive. The cluster thermal evolution is
simulated by replica exchange metadynamics (RE-META)
\cite{Metadynamics_method},
an approach which requires the simultaneous running of several molecular
dynamics simulations at different temperatures. Each simulation ({}``replica'')
is biased by an additional history-dependent \char`\"{}metadynamics
potential\char`\"{}\cite{laio_parrinello}. This artificial bias energy
term acts on an appropriately chosen small set of collective variables (CV),
global or local order parameter-like quantities representing some
physically motivated function of molecular coordinates. At fixed
time intervals, the replicas are forced to attempt an exchange of
their coordinate configurations. The exchange move is accepted or
rejected according to the usual Metropolis criterion\cite{metropolis, REMDS},
including the change of the history dependent CV-based bias potential
between the two replicas. The configurations visited in the RE-META
simulation correspond to visiting the free energy landscape of the
cluster -- as a function of the CV variables -- in parallel at all
temperatures. Combining replica exchange and metadynamics leads to
an overall kinetics which, even at low temperatures, rapidly becomes
diffusive in collective variable space, resulting in an extremely
efficient configuration sampling\cite{Metadynamics_method}. In RE-META, replica
exchange is improved because metadynamics explores high free energy
configurations within each replica, ultimately leading to faster decorrelation.
Metadynamics is in turn strengthened, since replica exchange improves
the sampling of the degrees of freedom that are not explicitly biased
-- thus mitigating one major  problem of metadynamics\cite{Metadynamics_method}.
The water nonamer example confirms that the free energies of the lowest
relevant states are indeed heavily affected by configurational entropies.
In particular, the entropy of the first excited state is higher
than that of the ground state, which causes for increasing temperature
an outright crossing of the two free energies and populations, a crossing
which does not occur in the vibrational 
{and rotational} approximation.

\section{The \ce{(H2O)9} cluster: RE-META calculations}

For the water nonamer \ce{(H2O)9}, we carried out RE-META simulations
with eleven replicas at the following temperatures: 25.8, 35.8, 47.8,
62.9, 82.4, 108.3, 136.4, 170.0, 205.5, 248.7 and 298.4 K. 
Since TIP4P bulk ice has a melting temperature of 
{232 K at 1 bar \cite{melting_point_TIP4P_2005},}
the higher temperatures correspond to a fast diffusing liquid-like regime,
the lower ones to a slow diffusing solid-like one. Our level of description
is entirely classical, and no quantum effects are included, either nuclear or electonic.
Because therefore our treatment ignores all quantum freezing and zero point
motion effects, the simulation results lose
their strict validity at very low and zero temperatures, a limit where they
should not be extrapolated.
{
The molecular dynamics simulation was performed in the NVT ensemble and
a cell volume of 18850 $\text{\AA}^3$, a volume where the probability of observing an isolated water  vapour molecule is extremely small.}
A N\'{o}se-Hoover chain thermostat was used
to control the temperature, the integration time step was set to 0.2
fs, and configurations were stored for analysis every ps. 
Each replica
was evolved for a total simulation time of 
{50 ns} 
and independently
biased by a metadynamics potential acting on a single collective variable
$S$ counting approximately the number of hydrogen bonds inside
the cluster:

\begin{equation}
S\left(r\right)=\sum_{ij}\frac{1-(r_{ij}/r_{0})^{8}}{1-(r_{ij}/r_{0})^{14}}.
\label{eq:CV}
\end{equation}
where the sum runs over proton-oxygen pair bonds belonging to different
molecules and $r_{0}=2$ \AA. The metadynamics potential has the form\cite{laio_parrinello}

\begin{equation}
V_{G}(S\left(r\right),t)=w\sum_{\begin{array}{c}
t'=\tau_{G},2\tau_{G},\cdots\\
t'<t\end{array}}\exp\left(-\frac{(S(r)-s(t'))^{2}}{2\delta
s^{2}}\right). \label{eq:V_G}
\end{equation}

where $s\left(t\right)=S\left(r\left(t\right)\right)$, $\delta s=0.15$ and 
{
$w=0.004$ meV.}
As customary in metadynamics, the collective variable  defined in  \ref{eq:CV}
is chosen on purely physical grounds, and is neither unique nor systematically
definable in a general case. However, while this arbitrariness could
be seen as endangering the reliability of metadynamics, the replica
exchange strategy does, as shown in Ref.\cite{Metadynamics_method}, effectively cure
the problem. In our calculation the exchange moves between configurations taken
at different temperatures were attempted every 2 ps, accepting the
moves with a
{
probability $\min(1,\exp\left(-\Delta\right) )$
with
\begin{equation} 
\Delta=\beta_iV_i(r_j)+\beta_jV_j(r_i)
-\beta_iV_i(r_i)-\beta_jV_j(r_j)
\end{equation}
where $i,j$ is the index of the replica, $V$ is the sum of the potential energy and the  bias,
$r$ are the coordinates and $\beta$ is the inverse temperature, which is different in each replica\cite{Metadynamics_method}.}
We found that after a typical equilibration time $\tau_{F}\sim 4$ ns the history-dependent
potential started growing evenly in CV space -- meaning independent of $S$ --
in all replicas, indicating that -$V_{G}(S,t)$ will eventually, for large $t$, provide a good
estimate of the free energy as a function of $S$\cite{laio_gervasio_09}.
After the equilibration time $\tau_{F}$ we could therefore
stop changing the history-dependent potential $V_{G}$ and
start collecting the statistics of occurrence of properly defined relevant states,
with the scope of computing their respective probability, and from
that their free energies and entropies in thermal equilibrium.

In order to analyze the results of the procedure it is necessary to
assign during the MD simulation each molecular coordinate configuration,
to the closest reference structure -- its relevant state. If all the
possible local minima of the potential energy surface were known,
one could classify the configurations according to their distance
from the reference structures, measured, for instance, by the appropriate
root mean square deviation of coordinates. This however would require a preliminary
search of all the minima. More importantly, in order to assign a configuration
to the correct relevant state, one should consider all permutations
between identical atoms, a procedure hugely cumbersome and computationally
expensive. We apply here a more conservative procedure that uses ideas
from the basin hopping technique\cite{wales_doye_1997}. 
Starting from all the coordinate
configurations generated by MD, we carried out a static potential
energy minimization
by conjugate gradients
evolving atomic coordinates
{
until the force on each atom was less than 0.0001 eV/\AA .} 
If a reference structure
with the same potential energy is already available, the instantaneous
MD configuration is assigned to that structure. Otherwise, a new reference
structure is introduced in the pool. 
{
Here two quenched structures are assumed to be the same if their energy difference
is smaller than 0.0001 eV.}
This procedure is still computationally
intensive, but allows an unambiguous assignment of all MD frames and
is insensitive to permutation of the atoms.

\begin{figure}
\includegraphics[clip,scale=0.4]{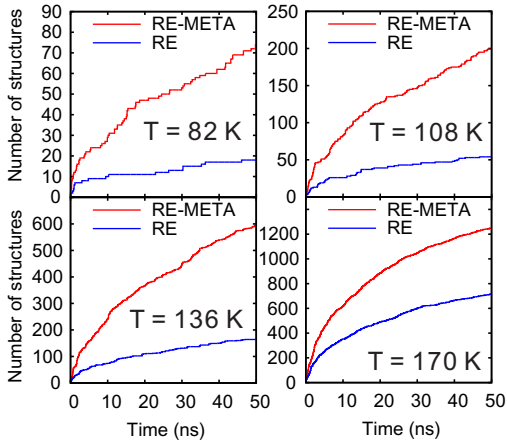}
\caption{The number of relevant states found as
a function of simulation time at different temperatures with simple
replica exchange (RE) and with RE-META. Note the great improvement
brought by addition of the history dependent metadynamics potential (
\ref{eq:V_G}), especially important at low temperature.}
\label{pstrnum}
\end{figure}

\ref{pstrnum} shows the number of independent structures that
are explored as a function of simulation time at different temperatures
for a normal replica exchange run, which we carried out in parallel for comparison,
and for RE-META. The number of structures visited is much larger in
the latter, a beneficial result of the metadynamic bias technique,
especially important at low temperature. This demostrates how RE-META
can be successfully used to carry out the first task normally undertaken
in cluster studies namely the search for all the low energy relevant
structures.

\begin{figure}
\includegraphics[clip,scale=0.15]{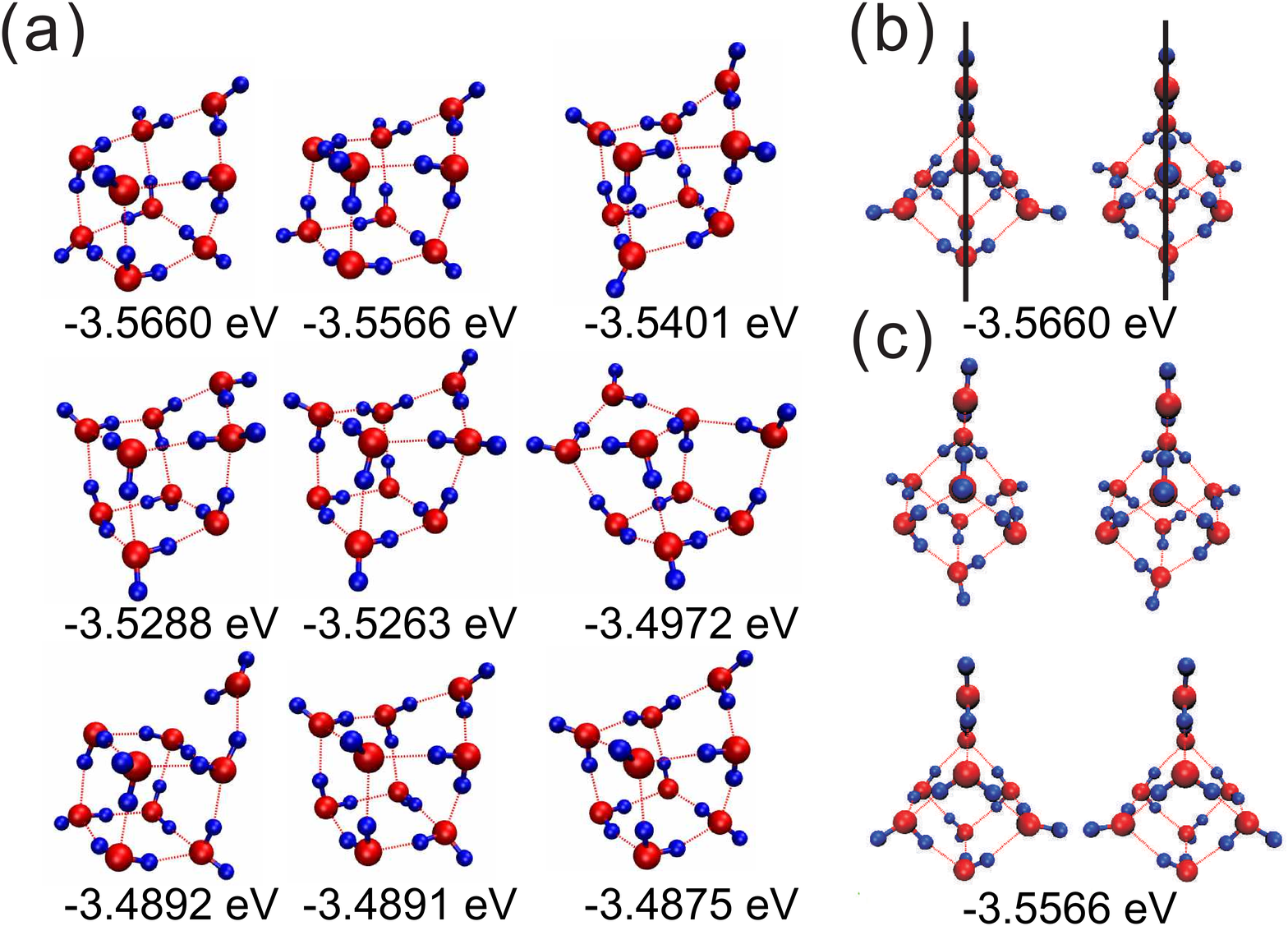}
\caption{(a)Lowest energy relevant states of the water nonamer. (b) Front and
back viewed configuration for state 0 (lowest energy state). Black line: symmetry plane.
(c) Front and back viewed  configurations for state 1(second lowest state).
For a fixed oxygen skeleton state 1 has twice larger multiplicity than state 0, which has
a mirror symmetry. Thus the configurational entropy of state
1 may be expected to exceed that of state 0 by approximately $\ln2$. }
\label{all_str}
\end{figure}

\ref{all_str}-a shows the molecular structure of the water nonamer
lowest energy relevant structures. The five lowest energy structures share the same
oxygen arrangement, and differ only by the position of the protons.
Eight oxygens form a regular 
{hexahedron} 
and an additional oxygen is
attached to one of the edges. The global minimum is identical to that
previously identified in ref. \cite{cluster_nonamer}. In all the
nine relevant structures 13 H-bonds are formed.

As seen in \ref{all_str} the lowest energy structure has a mirror
symmetry about the plane shown as a
black line in \ref{all_str}-b,
while the same mirror plane gives rise to two
distinguishable degenerate structures for the second lowest structure.
This, as we will see in the following, significantly influences their
respective configurational entropies.

\begin{figure}
\includegraphics[clip,scale=0.5]{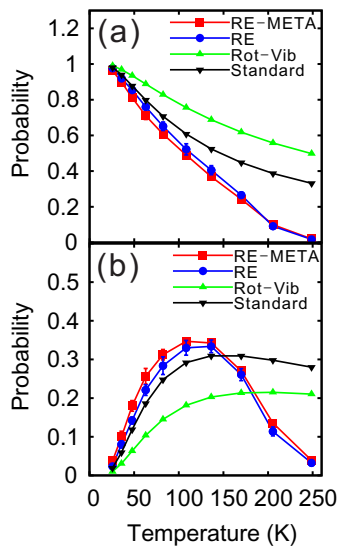}
\caption{{(a) occupancy probability for the
lowest energy relevant state (``0'') versus temperature,
with RE-META, normal RE,
rigid-rotor-harmonic approximation (``Rot-Vib'') and 
rigid-rotor-harmonic approximation with an additional  
the $\ln 2$ factor in the entropy of state 1 - state 8 (``Standard'').  
(b) occupancy
probability for the second lowest energy relevant state (``1''). The
statistical error bars in RE and RE-META represent the standard deviations
estimated by separately comparing averages in three successive portions of
the simulations. Note the substantial agreement between RE and RE-META 
and the large deviation of the rot-vib approximation,
showing its clear inadequacy. 
Note also that the occupancies of
states computed with RE and RE-META are fully consistent
with those computed using the standard approach 
at low temperature, 
and how the probability of state
1 overtakes that of state 0 around 150 K, as a result of its larger
configuration entropy.}}
\label{occ}
\end{figure}

We now come to our main point. Besides providing a tool for the
quick exploration of cluster structures, RE-META crucially yields an
accurate estimate of the probability to observe each relevant
structure $S$  as a function of temperature. Following Ref
\cite{10.1371/journal.pcbi.1000452}, we first compute the average bias potential
\begin{equation}
 V_{G}(S)=1/(\tau_{sim}-\tau_{F}) \int_{\tau_{F}}^{\tau_{sim}} dt V_{G}(S,t),
\label{eq:V_ave}
\end{equation}
where $\tau_{F}$ is  the equilibration time of
$V_{G}$, 4 ns in this work and $\tau_{sim}$ is the total simulation time. Denoting by $r_{i}$ all the
configurations generated by RE-META at temperature $T$, we estimate
the equilibrium probability to observe relevant structure $\alpha$
as
\begin{equation}
p^{\alpha}\left(T\right)=\frac{{\displaystyle \sum_{i\in\alpha}\exp\left({\normalcolor V}_{G}(S\left(r_{i}\right))/k_{B}T\right)}}{\sum_{\beta}\sum_{i\in\beta}\exp\left({\normalcolor V}_{G}(S\left(r_{i}\right)/k_{B}T)\right)}\label{eq:p_from_REM}
\end{equation}
where the notation $i\in\alpha$ means that the sum runs over the
configurations that are assigned to structure $\alpha$ as explained
above.
{
In order to assess the reliability of this approach, we also
computed the probability $p_{\alpha}$ in 
normal replica exchange  where the probability
is simply proportional to the number of times the structure $\alpha$
is observed.
As shown in \ref{occ}-a,b, the probabilities estimated
in the two manners are consistent. 
The small remaining discrepancy between RE and RE-META is due to statistical uncertainties. 
 We remark that the errors of methods based on statistical sampling 
are inevitably large at low T, as the relevant transitions are observed relatively rarely even if a powerful
enhanced sampling approach is used.}

\section{Comparison between RE-META and 
{rigid-rotor-harmonic vibration approximations}}

We calculate and compare in \ref{occ}-a,b the exact occupancy
probabilities of the two lowest relevant structures with the approximate
probabilities, now estimated using the {
rigid-rotor-harmonic vibration approximation} 
to the cluster vibrational spectrum of each state:

\begin{equation}
p^{\alpha}\left(T\right)=\frac{\exp(-F^{\alpha}/k_{B}T)}{\sum_{\beta}\exp(-F^{\beta}/k_{B}T)}\label{eq:p_from_boltz}
\end{equation}

where the {
rigid-rotor-harmonic free energies $F^{\alpha}$ are 
calculated as
\begin{equation}
F^{\alpha}=E^{\alpha}-kT\ln{q_{rot-vib}^{\alpha}}
\end{equation}
where $E^{\alpha}$ is the energy and
\begin{equation}
q_{rot-vib}^{\alpha}=q^{\alpha}_{rot}+q^{\alpha}_{vib}.
\end{equation}
\begin{equation}
q^{\alpha}_{rot}=
\left(
\frac{8\pi^3k_BT}
{h^2}
\right)^{3/2}
\frac
{\prod^3_{i=1}\sqrt{I^{\alpha}_i}}
{\sigma^{\alpha}\pi}
.
\end{equation}
\begin{equation}
q^{\alpha}_{vib}=
\prod_m
\frac{k_BT}
{h\omega^{\alpha}_m}.
\end{equation}
where $I^{\alpha}_i$ are the principal moments of inertia,
$\sigma^{\alpha}$ is the rotational symmetry number and
$\omega^{\alpha}_m$ is the vibrational frequency of mode $m$
obtained by diagonalizing
the Hessian matrix of the potential in structure $\alpha$ at
$T$=0. To keep the calculation simple, the summation in the denominator of  \ref{eq:p_from_boltz}
runs over the nine lowest energy states. 
We have checked
that if 30 structures are included in the summation, the probability
to observe all states between the 9th and 30th is only 0.2\% at 108K and
13\% at the highest temperature 250K.
It should  be noticed that the
rotational symmetry number of the lowest energy relevant states of the water nonamer are
all equal to 1 and the moments of inertia are similar. So, for this system, the rotational
entropy varies much less than the vibrational entropy. For example the rotational entropy
difference between state 0 and state 1 is only 7 \% of the vibrational entropy difference.
}
As is seen in \ref{occ}, the true occupancy probabilities from
RE-META and those predicted by the {
rigid-rotor-harmonic vibration approximation}
\ref{eq:p_from_boltz} differ very significantly.
For instance, at $T=200$
K,   \ref{eq:p_from_boltz} still predicts a significant
population for state 0 and state 1, while RE-META shows that the
population of state 0 has become negligible, while the population of
state 1 is only about 10 \%. 
{
Moreover, the  rot-vib  approximation
 \ref{eq:p_from_boltz} implies that state 0 is always more
populated than state 1, whereas according to RE-META the most
populated state at 150 K is state 1 and not state 0. }

\begin{figure}
\includegraphics[clip,scale=0.5]{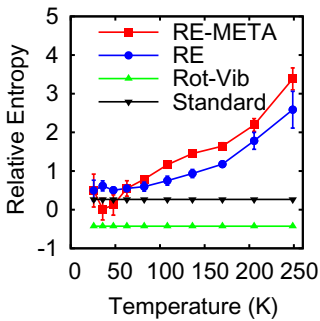}
\caption{
Entropy difference between state 1 and state 0 as a function of
T. Note the initially larger configurational entropy of state 1, further
growing with temperature, due to anharmonicity. Both effects are missed
by the {
rot-vib approximation, whereas the addition of $\ln(2)$ (the ``Standard'' approximation) provides a reliable result only well below 100 K.} }
\label{pentropy}
\end{figure}

{
We finally calculated the  entropy difference
between states 0 and 1
\begin{equation}
\Delta S=\frac{\Delta E-\Delta F}{k_B T}
\end{equation}
where $\Delta F$ and $\Delta E$ are the free energy and the internal energy differences.
\ref{pentropy} shows the
entropy difference estimated with RE-META,
RE, and 
rot-vib approximation}. This entropy difference is finite and
of order 1 even at the lower temperatures signalling an initial
configurational degeneracy of state 1, and grows further upon
heating.
The {
rot-vib approximation} on the other hand misses both effects.
Visual inspection of the structures assigned to relevant states 0 and 1 (data not shown)
reveals that up to T=200 K the oxygen skeleton remains largely unchanged.
Thus the main source of configurational entropy is proton disorder.

{The state 0 has a mirror symmetry, while state 1 does not.  The broken symmetry
implies that state 1 is two-fold degenerate.
Thus, the relative configurational entropy of
state 1 is roughly $\ln2$, which, at low temperature, approximately
agrees with the relatively large occupancy of state 1. 
Adding $\ln2$ to the rot-vib entropy for states (but not for state 0) leads to a positive ``Standard'' entropy difference (black curve in \ref{pentropy}), closer to the 
simulation results at small $T$.
At high temperature,
anharmonic contributions further favor state 1, making the entropy difference estimated even with the standard approach unreliable. 
Consistently, as shown in \ref{occ}, the occupancies estimated with the standard approach are compatible  with those estimated with RE-META at low $T$, but 
important deviations appear at $T>100 K$.
The comparison of the occupancies presented in  \ref{eq:p_from_boltz} and, even more, of the entropy differences of \ref{pentropy}, are the
central result of this work.
} 

\section{Discussion and conclusions}

\begin{figure}
\includegraphics[clip,scale=0.6]{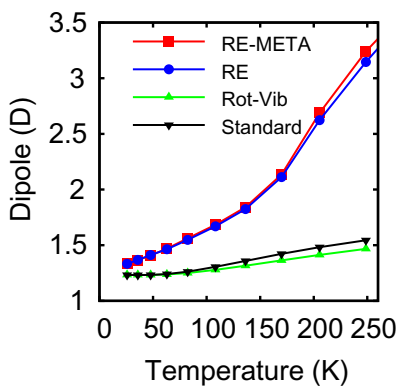}
\caption{Thermal averaged dipole moment of a water
nonamer, computed directly from the RE-META and RE trajectories and
using the probabilities ( \ref{eq:p_from_boltz}) estimated within the
{
``Rot-Vib'' and ``Standard'' approximation.}}
\label{fig:-Thermal-averaged}
\end{figure}

The cluster description obtained with RE-META can naturally be used
to predict measurable quantities. As an example, we computed the average
dipole of the water nonamer as a function of T. In \ref{fig:-Thermal-averaged},
the accurate RE-META computed dipole is compared with {
the rot-vib approximation}:
$D=\sum_{\alpha}D^{\alpha}p^{\alpha}\left(T\right)$ where $D^{\alpha}$
is the dipole in the reference structure $\alpha$ and the probabilities
$p_{\alpha}$ are estimated by  \ref{eq:p_from_boltz}. The exact
and the approximate results differ very significantly, underlining
the devastating effects of ignoring the configurational and anharmonic
entropies, especially at high $T$.

\begin{figure}
\includegraphics[clip,scale=0.6]{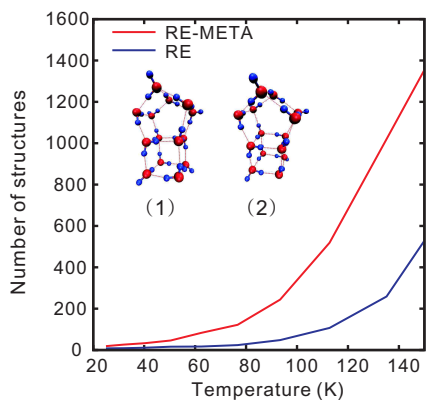}
\caption{The number of relevant states of \ce{(H2O)14} found as a function
of temperature with simple
replica exchange and with RE-META.
Improvement brought by RE-META method
is also notable. Insets: (1) Lowest relevant state.
(2) Second lowest relevant state.}
\label{fig:strnum14}
\end{figure}

Beyond the deliberately simple case study  \ce{(H2O)9} presented here,
RE-META method is applicable to larger water clusters, and of course to different systems,
once the collective variables are properly selected. The workload and complexity
grows considerably with size -- not surprisingly in view of the exponential growth
of phase space -- but the effectiveness is not impaired. As an example we considered
the larger water cluster \ce{(H2O)14}, hundreds of times more complex than the nonamer
{with a 10 ns simulation.}
\ref{fig:strnum14} shows the number of independent structures explored as a function of
simulation temperatures for RE and RE-META, underlining the strong superiority of the latter,
in particular its ability to unearth relevant structures that would otherwise be simply ignored.

In summary, we have shown how the full configurational complexity
of a small cluster can be directly, realistically and economically
accessed by a suitable sampling, combining replica exchange with metadynamics
techniques (RE-META). This proves extremely effective in searching
the cluster's relevant states, and in measuring their thermal population
as a function of T. In the water nonamer, these populations are shown
to differ largely from those routinely predicted by {
rot-vib approximation}. The crucial factor which influences the populations is the
configurational entropy of the relevant states, an entropy which is generally
inaccessible but which in RE-META is naturally and directly measured and
understood. The method holds a good promise of applicability in a great variety
of relevant systems, provided a set of good, physically motivated collective variables
can be identified.

\acknowledgement
ET acknowledges hospitality at Fudan, where this work began.  YZ acknowledges hospitality at SISSA,
where this work was mainly carried on.
Work at Fudan is partially supported by the National Science Foundation of China,
by special funds for major state basic research, by the research program of Shanghai
municipality and by MOE.  Work in Trieste is supported by the ESF EUROCORE FANAS project AFRI 
sponsored by CNR and by PRIN-COFIN 20087NX9Y7  contract of the Italian University and Research
Ministry.
\bibliography{water_cluster}
\newpage{}
\end{document}